\documentclass{webofc}
\usepackage[varg]{txfonts}   % Web of Conferences font
\usepackage{hyperref}
\usepackage{url}
\hypersetup{colorlinks=true,citecolor=blue,urlcolor=blue,linkcolor=blue}
\begin{document}
\title{Gamow shell model description of exceptional point in $^7$Li}

\author{\firstname{David} \lastname{Cardona Ochoa}\inst{1}\fnsep\thanks{\email{david.cardona@ganil.fr}} \and
        \firstname{Marek} \lastname{P{\l}oszajczak}\inst{1}\fnsep\thanks{\email{marek.ploszajczak@ganil.fr}} \and
        \firstname{Nicolas} \lastname{Michel}\inst{2,3}\fnsep\thanks{\email{nicolas.lj.michel@outlook.fr}}
        % etc.
}

\institute{Grand Acc\'el\'erateur National d'Ions Lourds (GANIL), CEA/DSM - CNRS/IN2P3,
BP 55027, F-14076 Caen Cedex, France 
\and
          CAS Key Laboratory of High Precision Nuclear Spectroscopy,Institute of Modern Physics, Chinese Academy of Sciences, Lanzhou 730000, China
\and
          School of Nuclear Science and Technology, University of Chinese Academy of Sciences, Beijing 100049, China
          }

\abstract{
We report the first identification of an exceptional point (EP) within the Gamow Shell Model in the Coupled Channels representation (GSM-CC). In the spectrum of $^{7}$Li, an EP is found for the $5/2^-$ doublet, where the two states coalesce in both energy and width, the phase rigidity vanishes, and the S-matrix develops a double pole. These features manifest directly in observables: the elastic cross section acquires a split-peak structure, and the phase shift shows a single $2\pi$ jump. This work demonstrates that GSM-CC provides a powerful framework to explore EP phenomena in nuclei and their experimental signatures.
}
\maketitle
\section{Introduction}
\label{intro}

A particularly striking feature of non-Hermitian systems is the emergence of exceptional points (EPs) \cite{Moiseyev_2011}, non-Hermitian degeneracies at which both eigenvalues and their associated eigenvectors coalesce. Unlike avoided level crossings in Hermitian systems, EPs occur at isolated points in parameter space where the Hamiltonian becomes non-diagonalizable.  Though long studied in mathematical literature \cite{Kato1995}, EPs manifest physically in open quantum systems (OQS) \cite{Rotter_2015}. They were first achieved experimentally in microwave cavities \cite{PhysRevLett.86.787}, but given the ubiquity of non-Hermitian systems, research on EPs spans a wide range of fields in physics such as optics \cite{PhysRevLett.108.173901}, atomic and molecular physics \cite{PhysRevLett.103.123003}, quantum phase transitions \cite{Heiss_2005} and even nuclear physics \cite{PhysRevC.80.034619} \cite{Okolowicz:2009mu}.

The OQS framework is particularly relevant in nuclear physics, where nuclei are not isolated but are embedded in the continuum of scattering states. As a result, many nuclear states acquire a finite lifetime through coupling to open decay channels. This continuum coupling gives rise to resonance phenomena and alters the structure of states near particle-emission thresholds, especially in weakly bound or unbound systems. In these regimes, decay dynamics play an essential role, and the mixing between discrete and continuum components becomes increasingly significant. A prominent theoretical tool that embodies this approach is the Gamow Shell Model (GSM) \cite{michel:in2p3-00331689}, which extends the traditional nuclear shell model by incorporating resonant and continuum single-particle states through a complex-energy Berggren basis coming from single poles of the S-matrix. Features of the double poles of the S-matrix in the realistic applications of this approach have not yet been studied. The GSM allows for a unified treatment of bound, resonant, and scattering states within the same formalism, providing a powerful framework to describe the interplay between nuclear structure and reaction dynamics. It has been applied to describe energy spectra, electromagnetic transitions, multipolar moments, elastic and inelastic p, n, $^2$H, $^3$H, $^3$He cross-sections, and p, n radiative capture cross sections.

 In the vicinity of an EP, the system exhibits nontrivial topology and heightened sensitivity to external perturbations, often manifesting as abrupt or non-analytic changes in physical observables \cite{refId0}  \cite{10.1063/1.4983809} \cite{Heiss_2012} \cite{A_I_Magunov_1999}. The aim of this work is to exploit the non-Hermitian nature of the GSM to find exceptional points in nuclear spectra to study the influences that these EPs have on different quantities related to $^7\text{Li}$.

\section{Theoretical Framework}\label{theory}

\subsection{GSM-CC}

The Gamow Shell Model in the Coupled Channels representation (GSM-CC) \cite{PhysRevC.99.044606} \cite{PhysRevC.89.034624} \cite{Michel2021} is a convenient formulation of the GSM that provides us with the necessary tools to study nuclear reaction and scattering properties. In this representation the A-body wave function is decomposed into binary reaction channels:

\begin{equation}
    |\Psi_{M}^J\rangle = \sum_c \int_0^\infty |(r,c)_{M}^J\rangle \frac{u_c^{JM}(r)}{r}r^2 dr ,
\end{equation}
where $r$ is the relative distance between the centers of mass of the projectile and target. $u_c^{JM}(r)$ is the radial amplitude describing the relative motion between the target and the projectile in the channel $c$ and is the solution to be determined by solving the GSM coupled-channel equations for every given total angular momentum $J$ and projection $M$. It should be emphasized that, in principle, the channel sum must include all possible decay modes: binary, ternary, and more complicated partitions. In practice, however, the expansion is truncated, and so far applications of GSM-CC have been restricted to binary channels only. The binary channel states $|(c,r)_{M}^J\rangle$ are defined as an antisymmetrized tensor product between the target state $|\Psi_T^{J_T}\rangle$ and the projectile state $|\Psi_P^{J_P}\rangle$ as follows :

\begin{equation}
 |(c,r)_{M}^J\rangle = \hat{\mathcal{A}}\left[ |\Psi_T^{J_T}\rangle \otimes |\Psi_P^{J_P}\rangle\right]_{M}^{J},
\end{equation}
where the channel index $c$ represents the mass partitions and their respective quantum numbers. The couplings between the angular momentum of the target $J_T$ and the projectile $J_P$ provides the total angular momentum $J$ of the system.
The coupled-channel equations can then be obtained from the Schrödinger equation $\hat{H}|\Psi_{M}^J\rangle = E |\Psi_{M}^J\rangle$, as:

\begin{equation}
    \sum_c \int_0^\infty r^2 \left(H_{cc'}(r, r') - E N_{cc'} (r,r')\right) \frac{u_c(r)}{r} =0,
\end{equation}
where $E$ is the scattering energy of the full system, and the kernels are defined as:

\begin{equation}
    H_{cc'}(r,r') = \langle(c,r)|\hat{H}|(c',r')\rangle
\end{equation}
\begin{equation}
    N_{cc'}(r,r') = \langle(c,r)|(c',r')\rangle.
\end{equation}

As the nucleons of both clusters interact via short-range interactions, the Hamiltonian $\hat{H}$ can be written as:

\begin{equation}
    \hat{H}=\hat{H}_T + \hat{H}_P + \hat{H}_{TP},
\end{equation}
where $\hat{H}_T$ is the intrinsic Hamiltonian of the target and $\hat{H}_P$ is the Hamiltonian of the projectile which is decomposed as $\hat{H}_{P}=\hat{H}_{int} + \hat{H}_{CM}$, where $\hat{H}_{CM}$ describes the movement of the center of mass of the projectile and $\hat{H}_{int}$ is its intrinsic Hamiltonian. Additionally, $\hat{H}_{TP}$ is the Hamiltonian that represents the interaction between clusters defined as: $\hat{H}_{TP} = \hat{H}-\hat{H}_{T}-\hat{H}_{P}$, in which $\hat{H}$ is the standard shell model Hamiltonian.

\subsection{Exceptional points in a 2x2 Matrix}

To describe some of the properties of EPs let us start with a simple model of two states $\epsilon_{1,2}(X)$ coupled to each other by some interaction $\omega(X)$ both of which depend on some set of variable parameters $X$. Represented by the following Hamiltonian:
\begin{equation}
H=
\begin{pmatrix}
\epsilon_1 & \omega \\
\omega & \epsilon_2
\end{pmatrix},  
\end{equation}
Finding the corresponding eigenvalues gives,

\begin{equation}
    E_{1,2} = \frac{\epsilon_1 + \epsilon_2}{2} \pm \frac{1}{2} \sqrt{(\epsilon_1 - \epsilon_2)^2 + 4 \omega^2}.
    \label{sqrt}
\end{equation}

In order to make the two eigenvalues equal, one needs to find a set of values for our parameters $X$ such that the square root in the previous expression is zero. Since the expression inside the square root is the sum of two squared quantities, $\epsilon_{1,2}$ and $\omega$ must be complex, rendering the Hamiltonian non-Hermitian.

In this non-Hermitian setting, the eigenvalues $E_{i}$ are complex and their associated eigenfunctions $|\psi_i\rangle$ become bi-orthogonal \cite{Brody_2014}. These wave functions, in the case of complex symmetric Hamiltonians, have to be normalized according to the bi-orthogonal inner product,
\begin{equation}
    \langle \psi_i^* | \psi_j \rangle = \delta_{ij},
\end{equation}
which replaces the standard Hermitian scalar product. This modified normalization reflects the fact that left and right eigenvectors form a dual basis, and that their overlap carries physical information about the non-Hermitian character of the system.  

A useful measure of this is the \textit{phase rigidity}, defined as
\begin{equation}
    r_i \equiv \frac{\langle \psi_i^*|\psi_i\rangle}{\langle \psi_i|\psi_i\rangle},
\end{equation}
which quantifies the degree to which the wave function retains a fixed phase. For Hermitian states one has $r_i=1$, reflecting a real wave function with a rigid phase. In contrast, close to an exceptional point the eigenfunctions coalesce and become self-orthogonal, leading to $r_i \to 0$. Thus the phase rigidity provides a direct indicator of how strongly non-Hermitian effects, such as resonance overlap and continuum coupling, affect the eigenstates.

    \subsection{S-matrix}
When investigating the role of EPs in decaying systems, it is useful to study the S-matrix. An EP appears as a double pole in the S-matrix. In the simplest case of a single resonance coupled to one decay channel, the S-matrix can be written as

\begin{equation}
    S=\frac{E-E_1+\frac{i}{2}\Gamma_1}{E-E_1-\frac{i}{2}\Gamma_1} = 1 + i\frac{\Gamma_1}{E-E_1-\frac{i}{2}\Gamma_1},
\end{equation}
where $E_1$ and $\Gamma_1$ are the energy and width of the resonance, respectively. If we extend this to describe two resonances, the S-matrix takes the form 

\begin{equation}
    S=\frac{(E-E_1+\frac{i}{2}\Gamma_1)(E-E_2+\frac{i}{2}\Gamma_2)}{(E-E_1-\frac{i}{2}\Gamma_1)(E-E_2-\frac{i}{2}\Gamma_2)}.
\end{equation}
At the EP, where both resonances have the same energies and widths, namely, $E_1=E_2\equiv E_{EP}$ and $\Gamma_1=\Gamma_2\equiv \Gamma_{EP}$, the S-matrix takes the form 

\begin{equation}
    S= 1 + 2i\frac{\Gamma_{EP}}{E-E_{EP}-\frac{i}{2}\Gamma_{EP}}- \frac{\Gamma_{EP}^2}{(E-E_{EP}-\frac{i}{2}\Gamma_{EP})^2},
    \label{smatrixep}
\end{equation}
and we obtain an expression that resembles the one for the single resonance case, plus a quadratic term which includes a double pole in the S-matrix. The interference between the linear and the quadratic terms has dramatic effects in quantities such as the elastic cross sections. Substituting \ref{smatrixep} into the standard relation for the elastic cross section in single-channel, spinless scattering, expressed in terms of the partial-wave S-matrix and restricting to the $l=0$ contribution,

\begin{equation}
  \sigma_{el}=  \frac{\pi}{k^2} \left| S - 1 \right|^2,
\end{equation}
one obtains 

\begin{equation}
     \sigma_{el}=  \frac{4\pi}{k^2} \frac{4(E-E_{EP})^2}{(E-E_{EP})^2-\frac{\Gamma_{EP}^2}{4}} \frac{\frac{\Gamma_{EP}^2}{4}}{(E-E_{EP})^2-\frac{\Gamma_{EP}^2}{4}},
     \label{equa}
\end{equation}
which resembles a Breit-Wigner resonance times an extra factor that goes to zero at the energy of the EP, leading to a split Lorentzian shape.

\section{Results}\label{results}

We will discuss the effects of the presence of an exceptional point on quantities associated to the $5/2^-$ doublet present in the spectrum of $^7\text{Li}$. The nucleus in question was modeled by $^4\text{He}$ core with 3 valence nucleons and using a channel basis comprising $[^4\text{He}(0_1^+) \otimes {}^3\text{H}(L_j)]^{J^\pi}$, $[^5\text{He}(K_i^\pi) \otimes {}^2\text{H}(L_j)]^{J^\pi}$ and $[^6\text{Li}(K_i^\pi) \otimes n(l_j)]^{J^\pi}$ channels. The cluster channels were constructed by coupling the partial waves $L_j = S_{1/2}$, $P_{1/2}$, $P_{3/2}$, $D_{3/2}$, $D_{5/2}$, $F_{5/2}$, $F_{7/2}$ of the wave function of ${}^3\text{H}$ with the inert core $^4\text{He}$ in the ground state $0^+$, and coupling the partial waves  $L_j = 3S_{1}$, $3P_{0}$, $3P_{1}$, $3P_{3}$,$3D_{1}$, $3D_{2}$, $3D_{3}$ of the wave function of $^2\text{H}$ with the $^5\text{He}$ target in the ground state $3/2^-$. The one-neutron channels were built by coupling the partial waves $l_j = s_{1/2}$, $p_{1/2}$, $p_{3/2}$, $d_{3/2}$, $d_{5/2}$, $f_{5/2}$, $f_{7/2}$ of the neutron wave functions with the states $K_i^\pi= 1_1^+$, $3_1^+$, $0_1^+$ $2_1^+$, $2_2^+$, $1_2^+$ of $^6\text{Li}$.

\begin{figure}[h!tb]
\centering
%\vspace*{1cm}       % Give the correct figure height in cm
\includegraphics[width=11cm,clip]{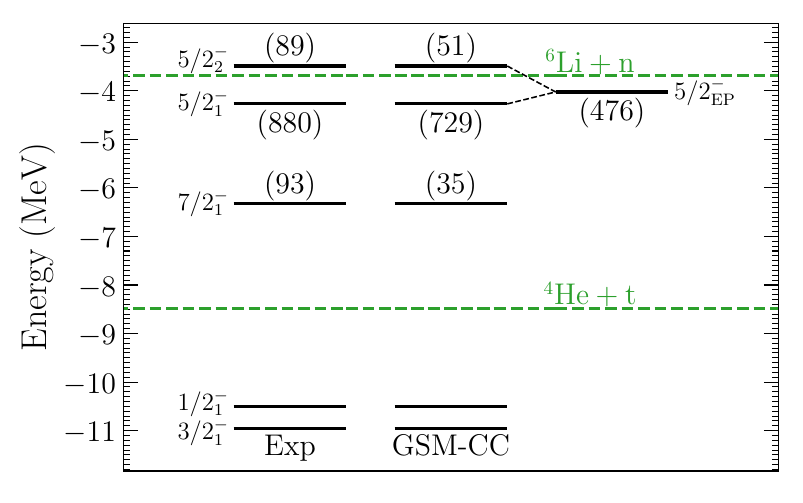}
\caption{Spectrum of $^7\text{Li}$ obtained in from the GSM-CC calculation showing the relevant particle emission thresholds, the energies are reproduced perfectly, with a slight underestimation of the experimental widths. On the right, the position of the found EP where the two $5/2^-$ states coalesce.}
\label{spec}       % Give a unique label
\end{figure}

The Hamiltonian consists of a one-body part of the Woods-Saxon (WS) type, plus a spin-orbit term and a Coulomb field, mimicking the core, and a nucleon-nucleon interaction in this case of the FHT type \cite{10.1143/PTP.62.981}. For this calculation the used WS and FHT parameters were taken from Linares Fernandez \textit{et al.} \cite{PhysRevC.108.044616}. Additionally, to account for the missing channels, the matrix elements for the the channel-channel couplings involving one-nucleon reaction channels have been re-scaled by corrective factor $c(J^\pi):$ $c(3/2^-)= 0.9972$,  $c(1/2^-)=1.0037$, $c(7/2^-)=1.0027$, $c(5/2^-)=0.96164$. Similarly, for the matrix elements involving cluster reaction channels we have for triton $c_{t}(5/2^-)=1.2693$ and deuteron $c_{d}(5/2^-)=1.3108$. 

Given this setup, Fig. \ref{spec} shows the obtained energy spectrum relative to the energy of the $^4\text{He}$ core, along with the relevant particle emission thresholds. The calculated GSM-CC energy spectrum is in perfect agreement with the experimental data, yet the widths are slightly smaller than the measured ones. 
We used the $l=1$ spin-orbit terms of the one-body potential for protons and neutrons as the variable parameters of the Hamiltonian to find the EP in the position shown in Fig.~\ref{spec}, just under the neutron threshold. 
\begin{figure*}[h!tb]
\centering
\includegraphics[width=6.4cm,clip]{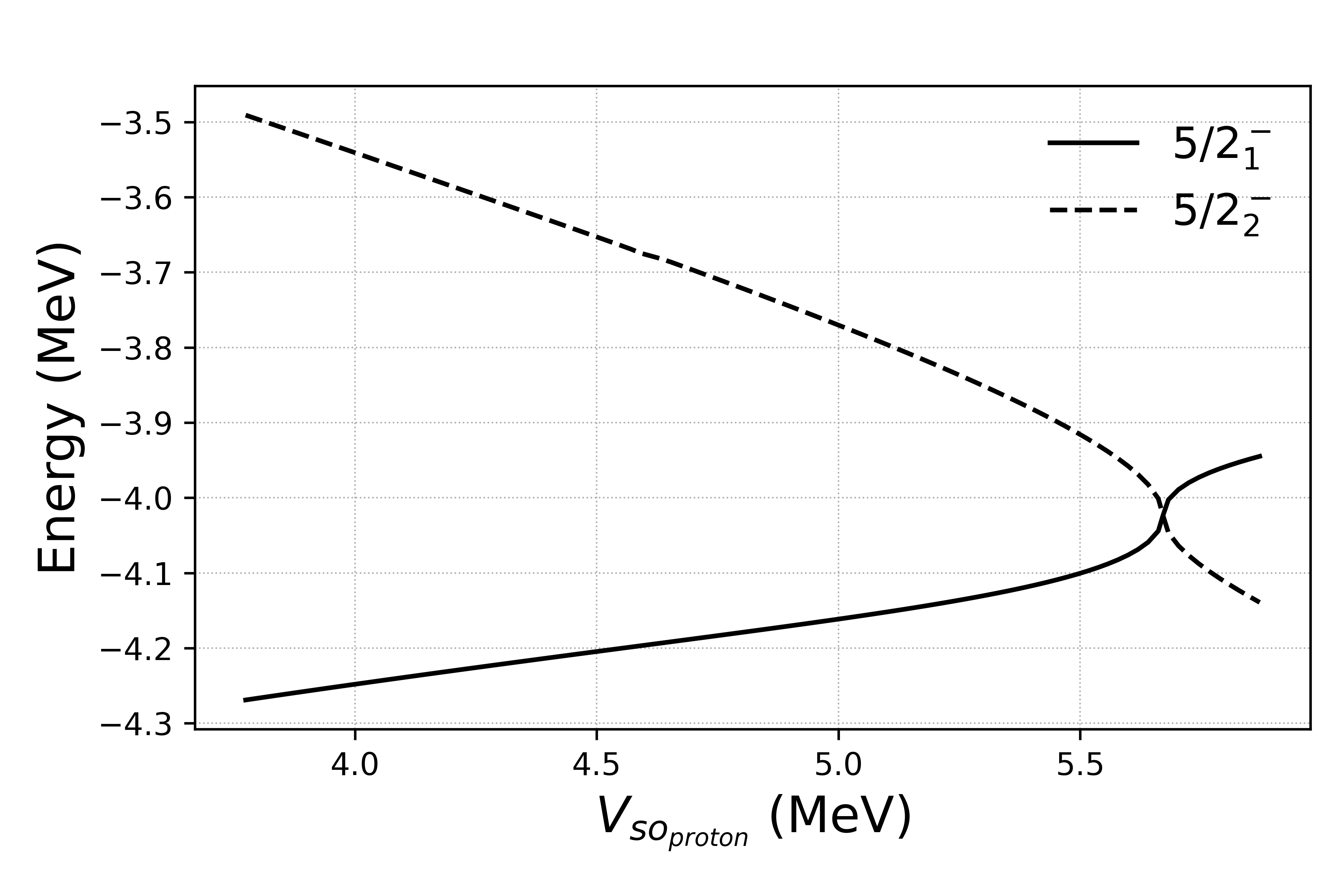}
\includegraphics[width=6.4cm,clip]{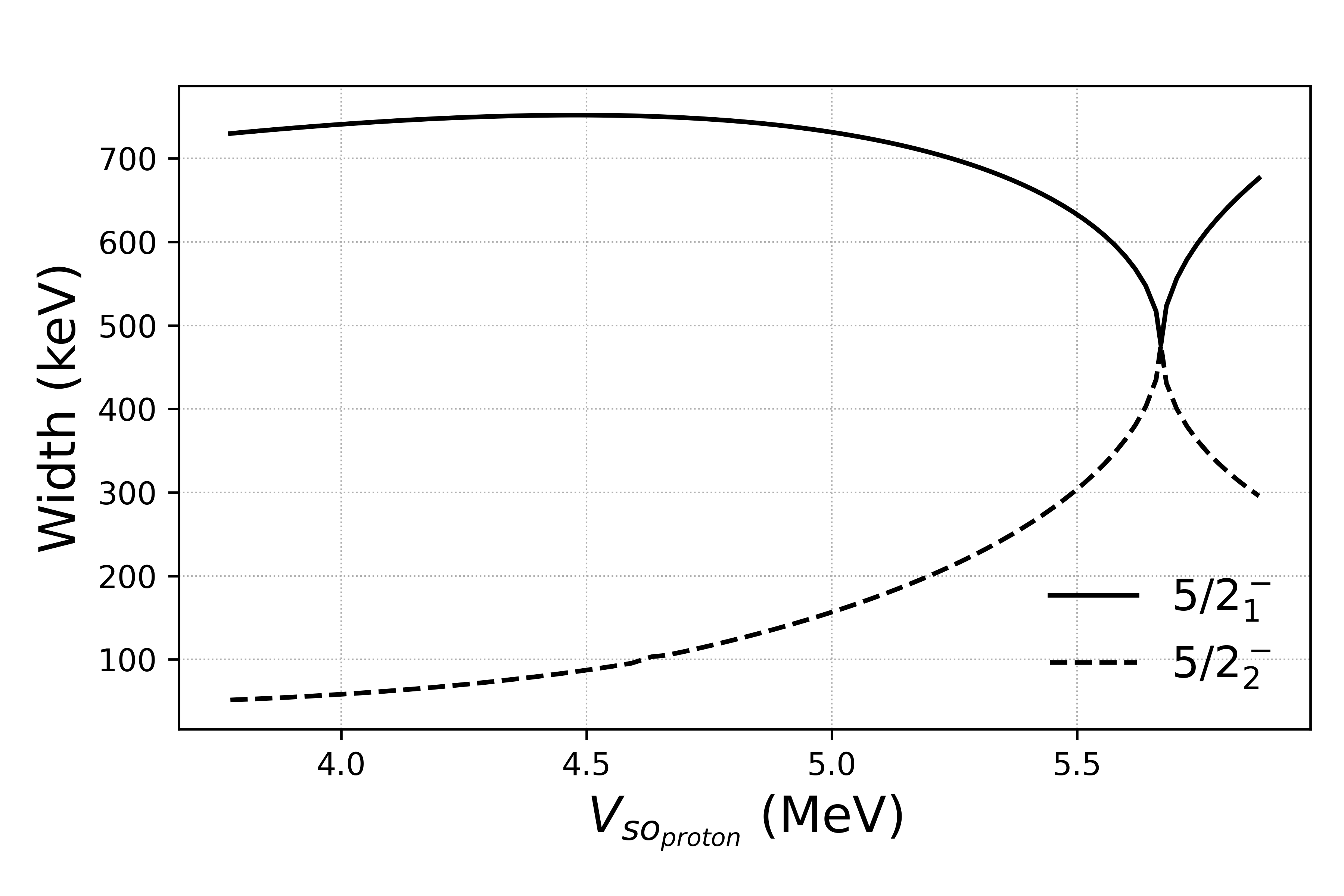}
\caption{Trajectories of the $5/2^-$ doublet towards the EP, exhibiting the characteristic square-root behavior of energies and widths."}
\label{ew}       % Give a unique label
\end{figure*}

Fig.~\ref{ew} shows the coalescence of the energies and widths of the $5/2^-$ doublet as a function of the control parameters, starting from the experimental fitted values $5/2^-_1:$ $E_1 = -4.269 \text{ MeV}$, $\Gamma_1=729 \text{ keV} $, and $5/2^-_2:$ $E_2 = -3.490 \text{ MeV}$, $\Gamma_2= 51.3 \text{ keV} $, and converging towards the found EP $5/2^-_{EP}:$ $E_{EP}= -4.023\text{ MeV}$, $\Gamma_{EP}=476\text{ keV}$, at $V_{so}(P)= 5.67 \text{ MeV}$ and $V_{so}(n)= 4.97 \text{ MeV}$. As described in \cite{Heiss_2012}, the square-root branch point structure of the eigenvalues, exemplified by Eq. (\ref{sqrt}),  implies that near the EP the trajectories of energies and widths show non-analytic behavior. This explains the strong parameter sensitivity visible in Fig.\ref{ew}, as even infinitesimal changes in the spin-orbit strengths lift the degeneracy and split the states with a characteristic square-root dependence, reflecting the topological nature of the EP.

\begin{figure*}[h!tb]
\centering

\sidecaption
\includegraphics[width=8cm,clip]{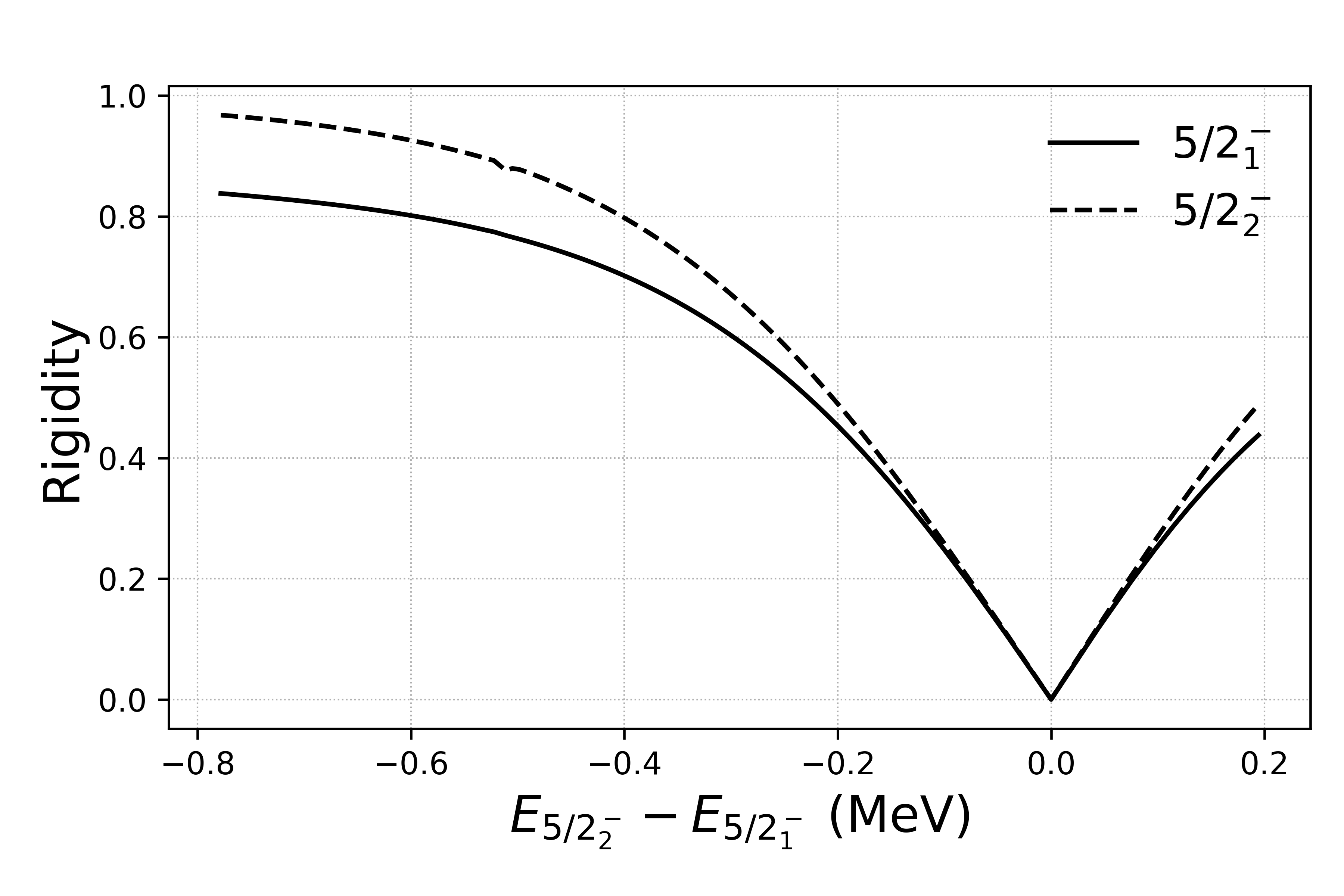}
\caption{Evolution of the rigidity of both $5/2^-$ states as a function of the separation in energy of both states, the rigidity reaches its minimum value of zero at the EP. }
\label{rig}       % Give a unique label
\end{figure*}
Fig. \ref{rig} shows the evolution of the phase rigidity of both $5/2^-$ states as a function of the separation energy between them. The initial deviation from unity indicates non-Hermitian effects already at the experimental values. As the states move closer, the rigidity decreases, reflecting their increasing non-orthogonality and stronger continuum admixture. At the exceptional point, the rigidity vanishes, signaling the coalescence of the eigenvalues, the self-orthogonality of the eigenfunctions, and the loss of distinction between the two states.

\begin{figure*}[h!tb]
\centering

\includegraphics[width=6.4cm,clip]{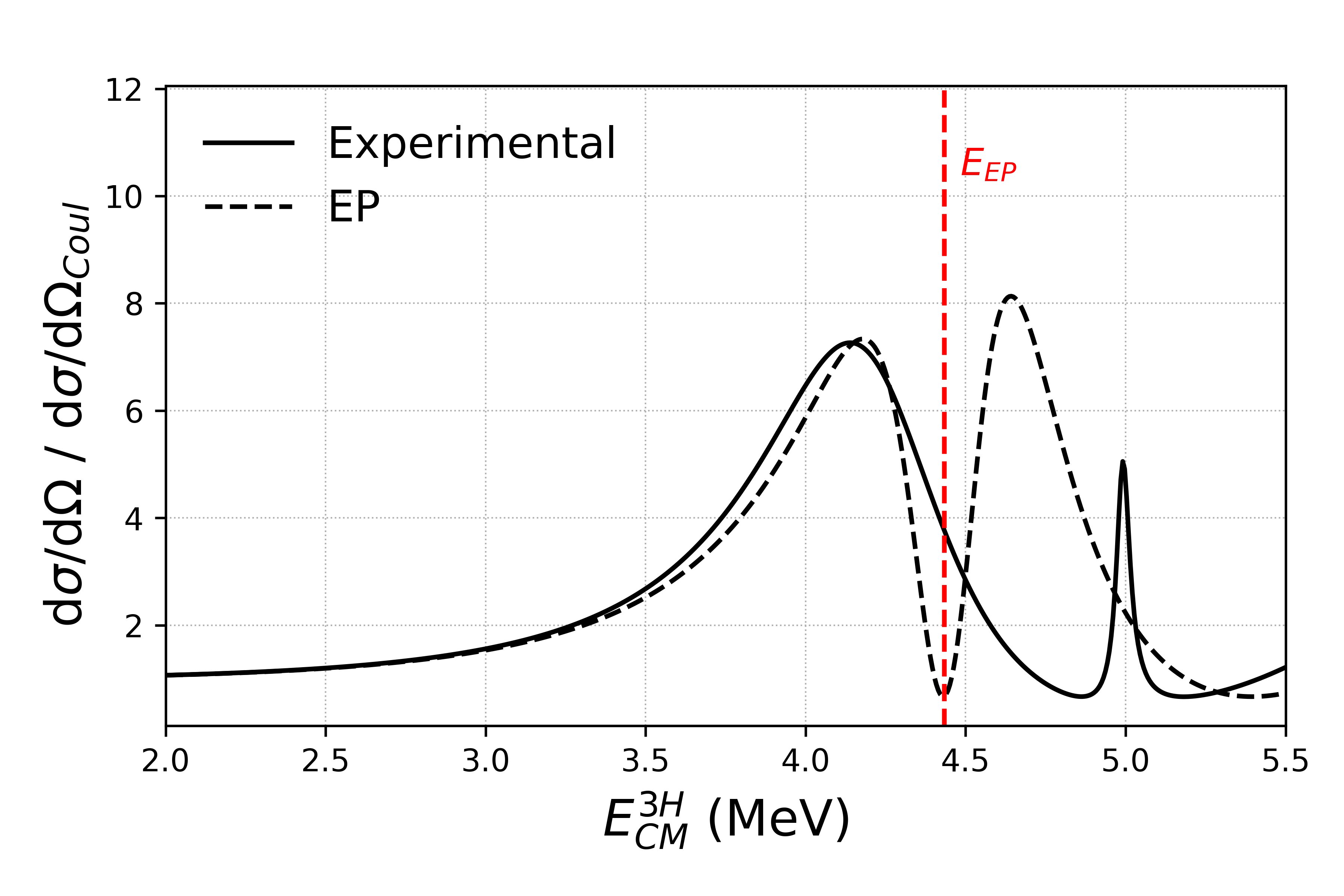}
\includegraphics[width=6.4cm,clip]{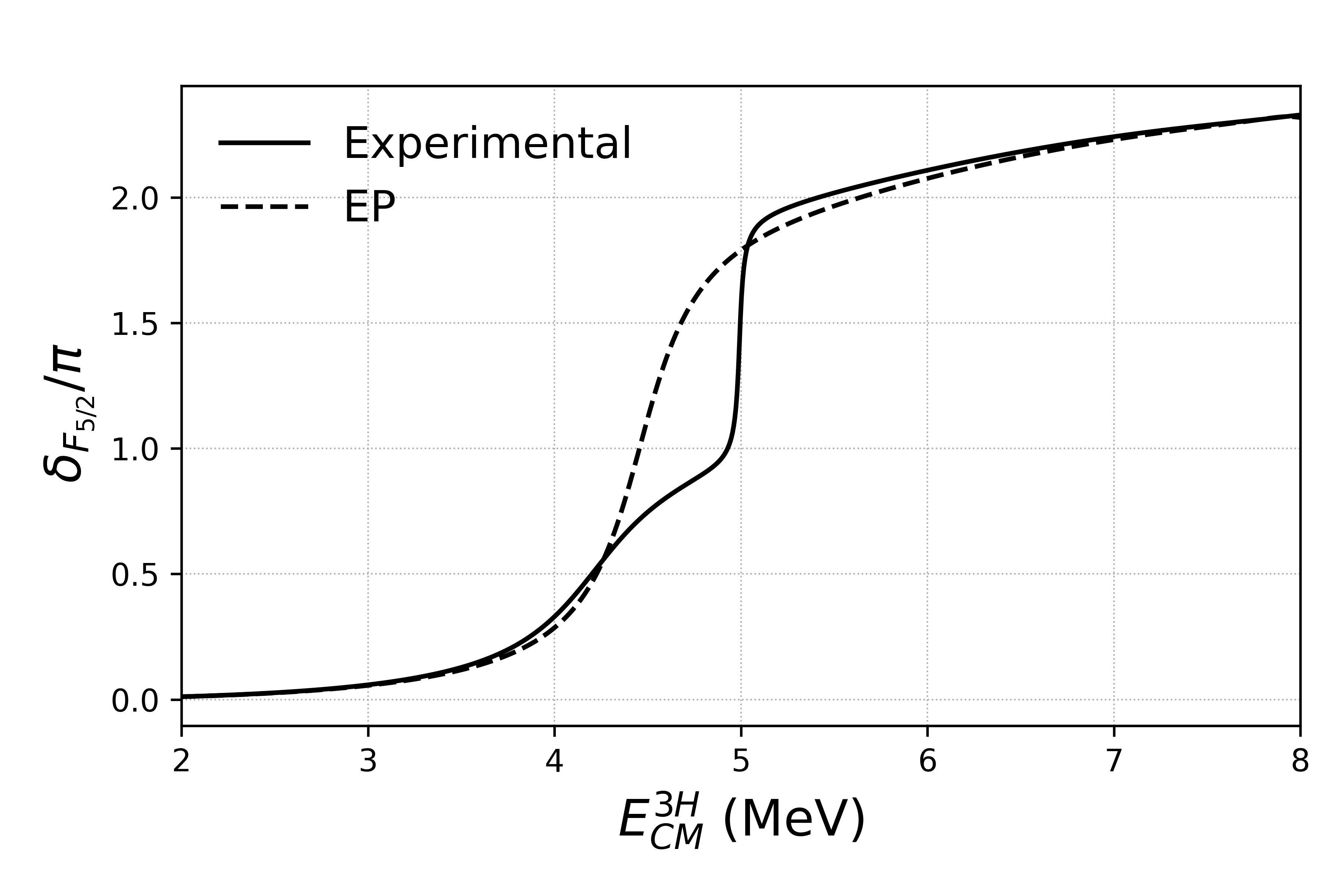}
\caption{On the left, the elastic cross section as a function of the center of mass energy for $^4\text{He}(^3\text{H},^3\text{H})^4\text{He}$ showing the characteristic split peak behavior at the energy of the state related to the EP. On the right, the phase shift of the $F_{5/2}$ partial wave of the $^3\text{H}$ projectile showing the $2\pi$ jump related to having a double pole in the S-matrix. }
\label{csps}       % Give a unique label
\end{figure*}
Now looking at reaction observables, the left side of Fig. \ref{csps} shows the contribution of the $5/2^-$ resonances to the elastic cross section for $^4\text{He}(^3\text{H},^3\text{H})^4\text{He}$, in which the solid line shows the behavior at the experimental fit and the dashed line shows the cross section at the found EP. The cross sections are plotted as a function of the center of mass energy. The Elastic cross section at the EP shows the split peak behavior predicted by the simplified model in equation \ref{equa} with the minimum in the position of the double pole of the S-matrix. 

The right side of Fig. \ref{csps} shows the phase shift for $^4\text{He}(^3\text{H},^3\text{H})^4\text{He}$ elastic scattering for the $F_{5/2}$ partial wave of $^3\text{H}$. The solid line shows the result for the experimental fit, for which we can see two distinct close to $\pi$ jumps as expected for the two resonances in the doublet, while the dashed line shows a single clear $2\pi$ jump which is another consequence of the presence of a double-pole in the S-matrix.

\section{Conclusions}

We have investigated the occurrence and impact of EPs in the spectrum of $^{7}$Li using the GSM-CC. By varying the proton and neutron spin--orbit terms of the one-body potential, an EP associated with the $5/2^-$ doublet was identified. At this point, the two states coalesce in energy and width, the phase rigidity collapses to zero, and the S-matrix develops a double pole. The consequences are clearly visible in reaction observables: the elastic cross section shows the predicted split-peak structure, while the phase shift exhibits a single $2\pi$ jump instead of two separate $\pi$ jumps.  

In nuclear physics, unlike in optical or atomic systems, the control parameters cannot be tuned experimentally to place the system exactly at the EP. Nevertheless, our results demonstrate that being in its proximity already leads to strong parameter sensitivity and non-trivial behavior of observables. This suggests that signatures of EPs may be accessible indirectly in nuclei through deviations in cross sections, phase shifts, and widths. Future work will extend these studies to other observables, such as electromagnetic transition strengths and spectroscopic factors, where EP effects are also expected to leave strong imprints.  

\vskip 0.3truecm
\textit{Acknowledgments --}
N. Michel wishes to thank GANIL for the hospitality where this work has been done.
This material is based upon work supported by
the National Natural Science Foundation of China under Grant No.12175281, and the State Key Laboratory of Nuclear Physics and Technology, Peking University under Grant No. NPT2020KFY13.

%
% BibTeX or Biber users please use (the style is already called in the class, ensure that the "woc.bst" style is in your local directory)
% \bibliography{your_bib_file} % Replace "your_bib_file" with the actual name of your .bib file
\bibliography{bib.bib}

\end{document}